\documentclass[aps,pre,floats,twocolumn,superscriptaddress]{revtex4}

\usepackage{epsfig}
\usepackage{latexsym,amsmath}

\begin{document}
\title{Conservation laws for voter-like models on directed networks}

\author{M. \'Angeles Serrano}
\affiliation{IFISC (CSIC-UIB) Instituto de F\'isica
Interdisciplinar y Sistemas Complejos, E07122 Palma de Mallorca,
Spain}

\author{Konstantin Klemm}
\affiliation{IFISC (CSIC-UIB) Instituto de F\'isica
Interdisciplinar y Sistemas Complejos, E07122 Palma de Mallorca,
Spain} \affiliation{Bioinformatics Group, Department of Computer
Science, University of Leipzig, Leipzig, Germany}

\author{Federico Vazquez}
\affiliation{IFISC (CSIC-UIB) Instituto de F\'isica
Interdisciplinar y Sistemas Complejos, E07122 Palma de Mallorca,
Spain}

\author{V\'ictor M. Egu\'iluz}
\affiliation{IFISC (CSIC-UIB) Instituto de F\'isica
Interdisciplinar y Sistemas Complejos, E07122 Palma de Mallorca,
Spain}

\author{Maxi San Miguel}
\affiliation{IFISC (CSIC-UIB) Instituto de F\'isica
Interdisciplinar y Sistemas Complejos, E07122 Palma de Mallorca,
Spain}

\date{\today}

\begin{abstract}
We study the voter model, under node and link update, and the
related invasion process on a single strongly connected component of a directed network. We
implement an analytical treatment in the thermodynamic limit using the
heterogeneous mean field assumption. From the dynamical rules at
the microscopic level, we find the  equations for the
evolution of the relative densities of nodes in a given state on
heterogeneous networks with arbitrary degree distribution and
degree-degree correlations. We prove that conserved quantities as
weighted linear superpositions of spin states exist for all
three processes and, for uncorrelated directed
networks, we derive their specific expressions. We also discuss the
time evolution of the relative densities that decay exponentially
to a homogeneous stationary value given by the conserved quantity. The conservation laws obtained in the thermodynamic limit for a system that does not order in that limit determine the probabilities of reaching the absorbing state for a finite system. The
contribution of each degree class to the conserved quantity is
determined by a local property. Depending on the dynamics, the
highest contribution is associated to influential nodes reaching a
large number of outgoing neighbors, not too influenceable ones
with a low number of incoming connections, or both at the same
time.
\end{abstract}

\pacs{02.50.-r,87.23.Ge,89.75.Fb}

\maketitle

\section{Introduction}

Conservation laws are intimately related to symmetries in the
systems they hold for. They play an important role in the characterization and
classification of different nonequilibrium processes of ordering dynamics. For example, in Kinetic
Ising models one distinguishes between Glauber (spin flip) and Kawasaki (spin exchange) dynamics.
Kawasaki dynamics fulfils a microscopic conservation law,
such that the total magnetization is conserved in each individual dynamical step of a stochastic realization.
This conservation law does not hold for Glauber
As a consequence, the Glauber and Kawasaki dynamics give
rise to different scaling laws for domain growth in coarsening processes~\cite{Gunton:1983}, and they define
different nonequilibrium universality classes.

In other types of nonequilibrium lattice models non-microscopic conservation laws are known to hold.
They are statistical conservation laws in which the conserved quantity
is an ensemble average defined over different
realizations of the stochastic dynamics for the same distribution
of initial conditions. Examples of such conservation laws occur for the voter
model~\cite{Clifford:1973,Holley:1975} or the invasion
process~\cite{Castellano:2005}. In particular, the role of the conservation law of the
magnetization and of the $Z_2$ symmetry (±1 states) in the voter dynamics universality class
has been studied in detail in the critical dimension d = 2 of regular lattices~\cite{Dornic:2001}.
The voter model is a paradigmatic model
of consensus dynamics in the social context~\cite{SanMiguel:2005,Castellano:2008} or, in the biological context, of competition of plant species in ecological communities~\cite{Chave:2001}.
In general, any Markov chain with at least two absorbing states reachable from all other
configurations has a conserved quantity when averaged over the ensemble.
Such a quantity determines the probability to eventually reach a
particular absorbing configuration in a finite system.
In some cases, this conservation law is of rather trivial nature as in the zero temperature Ising Glauber dynamics where
the magnetization sign is conserved. The voter model, the zero temperature Ising Glauber dynamics, and other related models of language
evolution~\cite{Castello:2006} or population
dynamics~\cite{Tilman:1997}, belong to the class of models with two absorbing states while epidemic
spreading dynamics, like the contact process~\cite{Boguna:2008a}
or the Susceptible-Infected-Susceptible
model~\cite{Romualdo:2001}, usually have a single absorbing state
with no conservation law.

While some of these questions have been studied for spin lattice models for a long time, conservation laws for dynamical processes on
complex networks~\cite{Albert:2002,Dorogovtsev:2003,Newman:2003,BarratBook:2008}
still remain a challenge. This issue has been considered for the voter
model~\cite{Clifford:1973,Holley:1975} or the invasion
process~\cite{Castellano:2005}
on undirected uncorrelated networks~\cite{Wu:2004,Suchecki:2005b,Sood:2005,VazquezF:2008}.
The link-update dynamics for the voter model has been found to
conserve the global magnetization~\cite{Suchecki:2005a}, while the
node update dynamics~\cite{Suchecki:2005a} and the invasion
process~\cite{Sood:2005} preserve a weighted global magnetization
where the contribution of each spin is calibrated by some function
of the degree of the corresponding node in the undirected network.
Such ensemble average conservation laws characterize
processes with two absorbing states accessible to the
dynamics, that compete to maintain an active state in the
thermodynamic limit. In finite networks, the conserved quantities
give the probabilities of reaching the uniform states and so act
as a bridge that enables some probabilistic predictive power of
the final dynamical state based on information about the initial
conditions. In addition, different finite size dynamical scaling properties can be related to different conservation laws~\cite{Suchecki:2005a}.

Much less has been done exploring dynamical processes on directed networks, with
the exception of the Ising model \cite{Sanchez:2002} and Boolean dynamics mainly
applied to biological problems~\cite{Kauffman:1969}. However, interactions
between pairs of elements are asymmetric in different systems including some
social networks~\cite{Newman:2001c}, where social ties are perceived or
implemented differently by the two individuals forming the connected pairs. Directed network representations rather than undirected ones become more
informative and adjusted to reality. In general, directed networks present
characteristic large-scale connectivity structures, the so-called bow-tie
architecture formed by a strongly connected component as a core structure and
peripheral in- and out-components~\cite{Broder:2000}. This organization, coupled
to the initial condition of the dynamics running on top, have an impact both on
the evolution of the processes and the final possible states of the
systems~\cite{Lieberman:2005,MinPark:2006,Jiang:2008}. In the voter model, leaf
nodes in the in-component never change their state thus sending an invariable
signal that can potentially propagate to the rest of the components of the
system. This is closely related to phenomena such as the presence of
zealots~\cite{Mobilia:2003,Mobilia:2007} in undirected networks. Both input or
output directional large-scale components and zealotry imply at the end an
external forcing on the dynamical processes that prevents reaching one of the absorbing states even for a finite network.
This is clearly illustrated by the
evolution of dynamical processes running on networks at the transition from a
pure strongly connected component to a complete bow-tie structure. In an
isolated and  strongly connected component, the voter dynamics keeps an active dynamical state in the thermodynamic limit,
but it leads to
a consensus (absorbing state) in a finite network as it happens on undirected networks. Thus, the appearance of an input
component in the large-scale structure of the network prevents the system from reaching an absorbing state for
random initial conditions~\cite{MinPark:2006}.

In this paper, we focus on dynamics of coupled two-state spin variables
and consider conserved quantities that are weighted sums of the spin
values. Specifically, we investigate the form of the conservation law for
the voter model --- under node and link update --- and the invasion
process in directed networks with arbitrary degree distribution and
degree-degree correlations. The directionality of the interactions is therefore
encoded in the topology. We restrict to a single strongly
connected component so that the absorbing state can be reached in a finite system, what seems realistic for a number of densely connected real networks like the world trade web~\cite{Serrano:2007b}. In Sec.~II, we present a detailed study of the node update version of the voter model and implement
an analytical treatment using the heterogeneous mean field assumption in the thermodynamic limit. From
the dynamical rules at the microscopic level, we find the  equations for
the evolution of the relative densities of nodes in one of the two possible states on heterogeneous
networks with arbitrary degree distribution and degree-degree correlations.
In this case, we prove that a conserved quantity as a weighted linear superposition of spin states exists.
In Sec.~III, we discuss the node-update voter model in
uncorrelated directed networks to derive analytical expression for the
conservation law and we also discuss the exponential decay of the relative
densities to their homogeneous stationary value, which is basically a function of the conserved quantity. We show how the conserved quantity determines the probability of reaching one of the two states in a finite network.
In Sec.~IV and V, we present the results of applying
the same methodology to the voter model with link update and the invasion
process, respectively. We conclude in Sec.~VI with a summary of results and open questions
for future research.

\section{The voter model on strongly connected components}
In the voter model under node update (VM), each node of a network
can exist in one of two possible states, $1$ or $0$~\footnote{We us this values $s=1,0$ in order to simplify computations instead of the usual spin notation $\sigma=\pm 1$. There is a direct mapping between both schemes $\sigma=2s-1$, and therefore for all the properties defined as a function of the states. For instance, the total magnetization $m$ in the $\{\pm 1\}$ scheme is related to the total magnetization $m'$ in the $\{0,1\}$ scheme through $m=2m'-1$.}. In a single dynamical event, a randomly selected node copies the state of one of its
neighbors, also selected at random. The link update dynamics of the Voter model selects instead a link~\cite{Suchecki:2005a}. Time is increased by $1/N$, so that the physical time is incremented by 1 after $N$ of such
events. On undirected networks, the node-update voter model conserves the
ensemble average of a weighted magnetization, where the
contribution of each spin is multiplied by the degree of the
corresponding node.

As defined above, the interactions in the voter dynamics are instantaneously
asymmetric since the updates always go in the same direction once the original
node is chosen independently of the undirectionality of the substrate. Hence,
the discussion of the voter model on directed networks comes out as a natural
one, where the directionality of the interaction is decoupled from the
dynamics and encoded in the structure of the substrate. The straightforward
generalization of the voter model on directed networks under node update
consists of selecting a node at random, and then assigning to it the state of
one of its incoming neighbors, also chosen at random. We will discuss this dynamics next in this section and Sec.~III, and the voter model with link update will be discussed later in Sec.~IV.

\subsection{Directed networks}
The topological structure of directed networks is more complex
than the one of undirected graphs. In purely directed
networks, without bidirectional links, the edges are
differentiated into incoming and outgoing, so that each vertex has
two coexisting degrees $k_\textrm{in}$ and $k_\textrm{out}$, with
total degree $k=k_\textrm{in}+k_\textrm{out}$. Hence, the degree
distribution for a directed network is a joint degree distribution
$P(k_\textrm{in},k_\textrm{out})\equiv P({\bf k})$ of in- and
out-degrees that in general may be correlated. We consider degree
correlations $P_\textrm{in}({\bf k'} | {\bf k})$ and
$P_\textrm{out}({\bf k'} | {\bf k})$, which respectively measure the
probability to reach a vertex of degree ${\bf k'}$ leaving from a
vertex of degree ${\bf k}$ using an incoming or outgoing edge of
the source vertex, and are related through the
following degree detailed balance condition~\cite{Boguna:2005}
\begin{equation}
k_\textrm{out} P({\bf k}) P_\textrm{out}({\bf k'} | {\bf k})=k'_\textrm{in} P({\bf k'}) P_\textrm{in}({\bf k} | {\bf
k'}).
\label{detailed_dir}
\end{equation}
This ensures that the network is closed and $\left< k_\textrm{in} \right>=\left< k_\textrm{out} \right>$.
Apart from the prescribed degrees and two point correlations, networks are maximally random.

At the macroscopic scale, the giant weakly connected component,
{\it i.e.}, the set of nodes that can communicate to each other
when considering the links as undirected
\cite{Molloy:1995,Molloy:1998,Havlin:2000,Callaway:2000,Newman:2001b},
becomes internally structured in three giant connected components, as well as other secondary structures such as
tubes or tendrils, forming a bow-tie architecture~\cite{Broder:2000}. The main component is the strongly connected component (SCC), a central core formed by the set of vertices that can be reached from each other following a directed path. The other two main components are peripheral components, the in component (IN) formed by all vertices from which the SCC is
reachable by a directed path but that cannot be reached from
there, and the out component (OUT) formed by all vertices that are
reachable from the SCC by a directed path but cannot reach the SCC
themselves. Percolation theory for purely directed networks was
first developed for uncorrelated
networks~\cite{Callaway:2000,Newman:2001b,Dorogovtsev:2001,Dorogovtsev:2001b,Serrano:2007c},
and directed random networks with arbitrary two point degree
correlations and bidirectional edges \cite{Boguna:2005}.

We restrict to networks forming a strongly connected component  without
peripheral components that would act on the SCC as sources of external
forcing. We will see that within the strongly connected component,
conservation laws preserve weighted magnetizations, where the weights are
dictated by the directed degrees.

\subsection{From microscopic dynamics to the drift equation under the heterogeneous mean field assumption}
To study the time evolution of the system, we
consider the drift part of the Langevin equation for the density of nodes in one of
the states from a microscopic description of the evolution of
single nodes' states applying a heterogeneous mean field approach. To our knowledge, this methodology, which
allows us to deal with dynamical processes on complex networks,
was first presented in Refs.~\cite{Catanzaro:2005b,Catanzaro:2008}
and recently used to study the contact process
\cite{Boguna:2008a}. In~\cite{VazquezF:2008b,VazquezF:2008}, a homogeneous mean field pair approximation was instead developed.

We focus on the microscopic state of nodes at some time $t$. Let
$s_u(t), u=1,...,N$, be a stochastic binary variable defined for
each of the $N$ nodes in the network which describes its state,
$0$ or $1$. The vector ${\bf s}(t) \equiv \{s_u(t)\},
u=1,\ldots,N$, completely defines the dynamical state of the
system at time $t$. Two more independent binary stochastic
variables $\mu(dt)$ and $\xi_u$ are defined  in order to model the
transitions between states of single nodes in an iteration.  After
a time interval $dt$, the variable $\mu(dt)$ for a given node $u$
takes the value $1$ or $0$ if $u$ was chosen or not,
respectively. In case node $u$ was selected, then $\xi_u$ assumes
the  value 1 [0] if $u$ copies a neighbor with state 1 [0]. We
assume that the occurrence of events in the voter dynamics follows
an independent Poisson process for each node, with constant rate
$\lambda$ for all of them, which corresponds to a Montecarlo step.
In the remainder we be set to $\lambda=1$ without loss of
generality. Thus, $\mu(dt)$ and $\xi_u$ have probability
distributions
\begin{equation}
P(\mu(dt))= dt \delta_{\mu(dt),1}+(1- dt) \delta_{\mu(dt),0},
\label{mu}
\end{equation}
\begin{equation}
P(\xi_u) = \Phi_u/k_{u,\textrm{in}} \delta_{\xi_u,1} +
(1-\Phi_u/k_{u,\textrm{in}})\delta_{\xi_u,0},
\label{xi}
\end{equation}
where $k_{u,\textrm{in}}$ is the incoming degree of node $u$, and
we have defined $\Phi_u(t)=\sum_v a_{vu}s_v(t)$. The adjacency
matrix $\{a_{vu}\}$ encodes the topological properties of the
directed network.  Element $a_{vu}$ has value one if there is a
directed link from $v$ to $u$ and zero otherwise, so that
$\Phi_u(t)$ stands for the number of state-one incoming neighbors
of node $u$ at time $t$. The matrix $\{a_{vu}\}$ is  symmetric for
undirected networks but for directed ones it is in general
asymmetric.

In terms of the above variables, the dynamical state $s_u(t)$ of node $u$
after an increment of time $dt$ is
\begin{equation}
s_u(t+dt)=\mu(dt)\xi_u + (1-\mu(dt))s_u(t). \label{micro1node}
\end{equation}
This equation, together with Eqs.~(\ref{mu}) and (\ref{xi}), give
the complete description of the evolution of the system, making
the formalism general and applicable to any network structure.

Although exact, this microscopic description is unmanageable. In
order to reduce the degrees of freedom, we apply a heterogeneous
mean-field hypothesis~\cite{Romualdo:2001} so that nodes with the
same degree ${\bf k}$ are assumed to be statistically independent and equivalent and can be aggregated in the same degree class
$\Upsilon({\bf k})\equiv \Upsilon(k_\textrm{in},k_\textrm{out})$.
Properties are then defined for each degree class, that will be
characterized by the relative density $m_{\bf k}(t)$, the ratio
between the number of state-one nodes within class $\Upsilon({\bf
k})$ and its number of nodes $N_{\bf k}$,
\begin{equation}
m_{\bf k}(t)=\frac{\sum_{u\epsilon\Upsilon({\bf k})}s_u(t)}{N_{\bf k}}.
\label{defrho}
\end{equation}
In the thermodynamic limit, the relative densities $m_{\bf k}(t)$ can be considered as continuous variables. Their time evolution can be
described by a Langevin equation~\cite{Gardiner:2004} with drift
and diffusion coefficients that are respectively given by the
first and second infinitesimal moments of the stochastic variables
$m_{\bf k}(t)$. Those moments can be derived from the microscopic
equation Eq.~(\ref{micro1node}) along with the definition in
Eq.~(\ref{defrho}). In the thermodynamic limit, it is possible to
prove that the diffusion term has a dependence
$1/\sqrt{N_k}$ on the system size as for undirected
networks~\cite{VazquezF:2008}, so that the drift term $A_k$ will
dominate. It is given by the average value over all possible
configurations of $m_{\bf k}(t+dt)$ conditioned to the state of
the system at time $t$,
\begin{equation}
\left< m_{\bf k}(t+dt)\right>_{m_{\bf k}(t)}=m_{\bf k}(t)+A_k(t)dt.
\end{equation}
From the microscopic dynamics
\begin{equation}
\left< s_u(t+dt) \right>_{{\bf s}(t)}=s_u(t) -dt \left[
s_u(t)-\Phi_u(t)/k_{u,\textrm{in}} \right],
\label{averagemicro}
\end{equation}
and summing this equation for all nodes in the degree class ${\bf
k}$ and dividing by the number of nodes $N_{\bf k}$, we arrive to
\begin{equation}
\left< m_{\bf k}(t+dt)\right>_{m_{\bf k}(t)}=m_{\bf k}(t) - dt
\left[m_{\bf k}(t)-\frac{1}{N_{\bf k}}\frac{1}{k_\textrm{in}
}\sum_{u\epsilon\Upsilon({\bf k})} \Phi_u(t) \right], \label{averagemeso}
\end{equation}
and from here to
\begin{equation}
A_k(t)=-m_{\bf k}(t)+\frac{1}{N_{\bf k}}\frac{1}{k_\textrm{in}
}\sum_{u\epsilon\Upsilon({\bf k})} \Phi_u(t).
\label{driftermV}
\end{equation}
The adjacency matrix contained in
$\Phi_u(t)$ can be coarse-grained as well, so that a differential
equation for the relative densities can eventually be written.
This coarse-graining restricts the
validity of the equations to random complex networks (and not
lattices), since we assume all nodes in the same degree class to
be statistically independent. With these assumptions,
\begin{eqnarray}
\sum_{u\epsilon\Upsilon({\bf k})} \Phi_u(t)&=&\sum_{\bf
k'}\sum_{v\epsilon\Upsilon({\bf k'})}\sum_{u\epsilon\Upsilon({\bf k})}
a_{vu}s_v(t)\nonumber \\ &=& \sum_{\bf k'}\bar{a}_{\bf k'k}N_{\bf k} N_{\bf
k'}m_{\bf k'}(t).
\end{eqnarray}
At this point, we restrict to directed networks organized at the large scale into a SCC without IN and OUT. This allows us to
write
\begin{equation}
\bar{a}_{\bf k'k}=\frac{E_{\bf k'k}}{N_{\bf k} N_{\bf k'}}=\frac{k'_\textrm{out}
P_\textrm{out}({\bf k}|{\bf k'})}{N_{\bf k}}=\frac{k_\textrm{in}P_\textrm{in}({\bf k'}|{\bf k})}{N_{\bf k'}},
\end{equation}
where $E_{\bf k'k}$ is the asymmetric matrix of the number of connections from
the class of vertices of degree ${\bf k'}$ to the class of vertices of degree
${\bf k}$, and where we have made use of the detailed balance condition
Eq.~(\ref{detailed_dir}).

Inserting these results into Eq.~(\ref{driftermV}), we arrive to
the equation for the evolution of the relative density in the
degree class ${\bf k}$ of a purely directed correlated network
(disregarding diffusion terms),
\begin{equation}
\frac{d m_{\bf k}(t)}{dt}=-m_{\bf k}(t)+ \sum_{\bf k'} P_\textrm{in}({\bf k'}|{\bf k})
m_{\bf k'}(t).
\label{driftV}
\end{equation}
Let us recall that this result is valid for the ensemble of
networks defined by the degree distribution $P({\bf k})$ and the
degree correlations $P_\textrm{in}({\bf k'} | {\bf k})$ and
$P_\textrm{out}({\bf k'} | {\bf k})$, but otherwise maximally
random. Notice that big enough networks present good statistical
quality at the level of degree classes and are also well described
by this equation. Finally, in the thermodynamic limit, the
Langevin equation loses its noise term because of the dependence
on the system size and reduces to Eq.~(\ref{driftV}), so that
$m_{\bf k}(t)$ becomes a deterministic variable. Nevertheless,
since the process is linear, Eq.~(\ref{driftV}) is always valid
even for finite systems understanding that in this case the
variables are averages over realizations of the process with the
same distribution of initial conditions.

\subsection{Conserved quantity on directed networks with degree-degree correlations}
For correlated networks, $m_{\bf k}(t)=\sum_{\bf k'}
P_\textrm{in}({\bf k'}|{\bf k}) m_{\bf k'}(t)$ in the stationary
state and hence all relative densities are entangled through
topological correlations. This equation corresponds indeed to an
eigenvector problem, since $\{m_{\bf k}(t)\}$ can be thought as the
eigenvector of the matrix $\{P_\textrm{in}({\bf k'}|{\bf k})\}$
with eigenvalue one.

We prove next that, within the heterogeneous mean
field approach and for the correlated directed networks we are
considering, there is a conserved quantity given as a linear
superposition of the form $\omega=\sum_{\bf k} \varphi_{\bf k}
m_{\bf k}(t)$. From Eq.~(\ref{driftV}), its evolution is given by
\begin{equation}
\frac{d \omega}{dt}=-\omega+ \sum_{\bf k}\sum_{\bf k'} \varphi_{\bf k}
P_\textrm{in}({\bf k'}|{\bf k}) m_{\bf k'}(t),
\end{equation}
and imposing that $d\omega/dt=0$, we obtain
\begin{equation}
\sum_{\bf k}\varphi_{\bf k}m_{\bf k}(t)=\sum_{\bf k}\sum_{\bf k'}  P_\textrm{in}({\bf
k}|{\bf k'}) \varphi_{\bf k'}m_{\bf k}(t). \label{cq}
\end{equation}
For each density
\begin{equation}
\varphi_{\bf k}=\sum_{\bf k'}  P_\textrm{in}({\bf k}|{\bf k'}) \varphi_{\bf k'}.
\label{varphiV}
\end{equation}
This is an eigenvector equation that has a solution if the matrix
$\{P_\textrm{in}({\bf k}|{\bf k'})\}$ has an eigenvalue equal to
one with $\{\varphi_{\bf k}\}$ the corresponding eigenvector. One
can prove that this eigenvector with eigenvalue one exists by
summing both sides of the previous equation over ${\bf k}$. Using
the normalization of the conditional probability $\sum_{\bf k}
P_\textrm{in}({\bf k}|{\bf k'})=1$, one eventually arrives to a
trivial identity~\footnote{For a wider validity range, the same can
be proved at the microscopic level from equation
Eq.~(\ref{micro1node}), which is exact for any graph, with the
only assumption that the adjacency matrix represents a SCC. One
has to assume $\omega=\sum_{u} c_u s_u$, but the procedure is the
same.}. The fact that the coefficients $\varphi_{\bf k}$ that
modulate the contributions of the different $m_{\bf k}$ to the
conserved weighted magnetization correspond to the entries of the
eigenvector of a certain characteristic matrix with eigenvalue one
also applies to other similar dynamical processes, such as the
link dynamics and the invasion process, as we will show.

This proves that a conserved quantity of the form of a linear functional
exists but, in general, it is not
possible to derive its value without further specifying the form of the
degree-degree correlations in the network.

\section{Voter model on uncorrelated SCCs}
When two-point correlations are absent, the transition probabilities become
independent of the degree of the source vertex. In this situation,
\begin{equation}
 P_\textrm{out}({\bf k'} | {\bf k})=\frac{k'_\textrm{in} P({\bf k'})}{\langle k_\textrm{in} \rangle} \mbox{
 , } P_\textrm{in}({\bf k'} | {\bf k})=\frac{k'_\textrm{out} P({\bf k'})}{\langle k_\textrm{in} \rangle},
\label{transitioninout_uncorrelated}
\end{equation}
and using these expressions, Eq.~(\ref{driftV}) becomes
\begin{equation}
\frac{d m_{\bf k}(t)}{dt}=-m_{\bf k}(t)+ \omega_\textrm{out},
\label{driftVunc}
\end{equation}
where we have defined
\begin{equation}
\omega_\textrm{out}=\frac{1}{\left< k_\textrm{in} \right>}\sum_{\bf k} k_\textrm{out}P({\bf k}) m_{\bf k}(t).
\label{wo}
\end{equation}
Therefore, in the
stationary state $m_{\bf k}=\omega_\textrm{out} \mbox{  } \forall
{\bf k}$ and $\omega_\textrm{out}$ is a conserved quantity in uncorrelated networks, which immediately follows from Eq.~(\ref{driftVunc}). In general,
it is not preserved in strongly connected
components of directed networks with degree-degree correlations. This is in contrast to undirected networks, where
the conserved quantity $\omega=\left( \sum_{k} kP(k)m_{k}(t) \right) / \langle k
\rangle$ is preserved even in the correlated case and indeed for
any structure~\cite{Suchecki:2005a}. Going back to the uncorrelated case, notice that the out-degree is the quantity that
weights the contribution of the nodes to the conserved quantity. From a local perspective, what seems therefore important in the VM is
to be able to influence a large number of partners

In uncorrelated networks, the convergence of the state-one
relative densities to their stationary value can be easily computed. From Eq.~(\ref{driftVunc}), taking into account
that $\omega_\textrm{out}$ is a conserved quantity and for a given
initial condition $m_{\bf k}(0)$, it is straightforward to arrive
to the solution
\begin{equation}
m_{\bf k}(t)=\omega_\textrm{out} + \left( m_{\bf
k}(0)- \omega_\textrm{out} \right) e^{-t},
\label{decayV}
\end{equation}
where we have substituted $\left <k_\textrm{in} \right>$ by $\left
<k_\textrm{out} \right>$.
Thus, all the densities decay exponentially fast to the stationary
value $m_{\bf k}^{st} = \omega_\textrm{out}$ and the relaxation
time is for all of them equal and independent of the degrees.

In the thermodynamic limit, the partially ordered stationary state
is stable, while finite-size fluctuations eventually bring the
system to one of the two possible unanimity states. The probability $P_1$ that the system ends up with all nodes in state
one ($m_{\bf k}=1, \forall {\bf k}$) is given by the initial condition, that fixes the value of the conserved quantity at the beginning of the process. To see this, one takes into account that $\omega_\textrm{out}$ is an ensemble average conserved quantity of the form in Eq.~(\ref{wo}), from which
\begin{equation}
\omega_\textrm{out}=P_1.
\end{equation}
This is in agreement with the fact that, in general, the Markov property of a stochastic process, if present, trivially ensures that the exit probability is a conserved quantity corresponding to a time-translation invariance. If the process has one absorbing state, the exit probability has a constant value one but, if the process has two or more absorbing barriers, the probability of reaching one of those is not trivial any more.

It is also interesting to investigate what happens to the quantity
$\upsilon_{i}(t)=\left( \sum_{\bf k} k_\textrm{in}P({\bf k})
m_{\bf k}(t)\right) / \langle k_\textrm{in} \rangle$, which involves in-degree
instead of out-degree. In the uncorrelated case, and disregarding
fluctuations,
$\upsilon_{i}(t)=(\upsilon_{i}(0)-\omega_\textrm{out})e^{-t}+\omega_\textrm{out}$,
that is, in general $\upsilon_{i}$ decays exponentially fast to
$\omega_\textrm{out}$. The quantity $\upsilon_{i}(0)$ depends on
the initial condition. If this is homogeneous over degree classes,
then $\upsilon_{i}(0)=\omega_\textrm{out}$ and $\upsilon_{i}(t)$
remains constant.

In order to check the convergence of the sate one relative
densities to the conserved quantity, we have run numerical
simulations of the voter model dynamics on a random uncorrelated network of
size $N=10^5$, scale-free in-degree distribution with exponent
$2.5$ and exponential out-degree distribution.  To obtain an
initial state that is inhomogeneous in the densities $m_{\bf k}$,
we have chosen an initial configuration in which half of  the
nodes with the lowest out-degree have state zero, and the other
half have  state one.  In this way, initial densities $m_{\bf
k}(0)$ in classes with $k_\textrm{out}$ lower than $4$ were small
or zero, while densities in classes with $k_\textrm{out}$ larger
than $4$ were one.

\begin{figure}[t]
\begin{center}
\includegraphics[width=3.4in, clip=true]{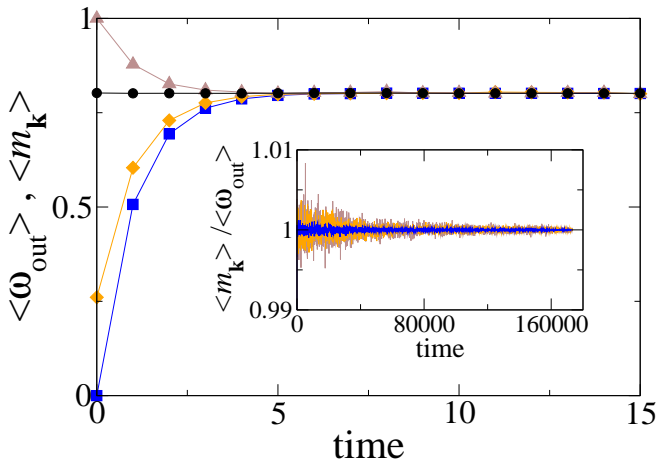}
\end{center}
\caption{Time evolution of the conserved quantity  $\omega_\textrm{out}$
(circles) and the densities of state-one nodes $m_{\bf k}$ in  degree classes
${\bf k}=(k_\textrm{in},k_\textrm{out})=(2,1)$ (squares), $(4,3)$ (diamonds)
and $(3,9)$ (triangles), for the voter model dynamics. Curves correspond to
averages over $100$ realizations on a single random uncorrelated network with $N=10^5$
nodes, scale-free in-degree distribution with exponent $2.5$ and exponential
out-degree distribution. While $\left< \omega_\textrm{out} \right>$ remains
roughly constant over time, the densities quickly decay to the stationary
value $\left< \omega_\textrm{out} \right>$.  The  inset shows that the ratio
between the densities and the conserved quantity is close to one during the
entire  evolution.}
\label{Fig1}
\end{figure}

\begin{figure}[t]
\begin{center}
\includegraphics[width=3.0in, clip=true]{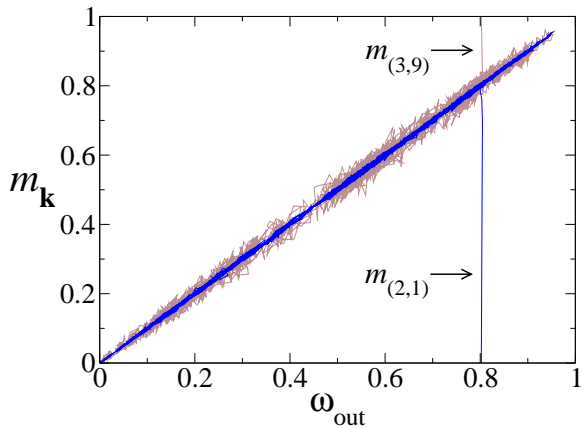}
\end{center}
\caption{Densities of state-one nodes $m_{(2,1)}$ and $m_{(3,9)}$ vs
$\omega_\textrm{out}$ in  a single realization of the voter model dynamics on
the same network of Fig.~\ref{Fig1}.  The trajectories of classes $(2,1)$ and
$(3,9)$ start at the positions $(0.8,0)$  and $(0.8,1.0)$ respectively, then
they quickly hit and move along the diagonal  $m_{\bf k}=\omega_\textrm{out}$,
until they reach the  zero-state consensus point $m_{(2,1)}=m_{(3,9)}=0$.}
\label{Fig2}
\end{figure}

In Fig.~\ref{Fig1}, we plot the average of the conserved quantity
$\omega_\textrm{out}$ and the densities for classes  ${\bf
k}=(k_\textrm{in},k_\textrm{out})=(2,1)$, $(4,3)$ and $(3,9)$ vs
time, over $100$ independent  realizations starting from the same
initial condition as mentioned above.  As predicted by the theory,
we observe that $\left< \omega_\textrm{out} \right>$ stays
constant over time, whereas the three densities converge  to the
average of the stationary value $m_{\bf k}^{st}$, in a time of
order $10$.  We note that, apart from finite size fluctuations,
the convergence of the  densities to $m_{\bf k}^{st}$ happens for every
realization. This can be seen in Fig.~\ref{Fig2}, where we show
the evolution of  $m_{(2,1)}$ and $m_{(3,9)}$ vs $\omega_\textrm{out}$
in a single run. After a short transient, the densities and the
conserved quantity start to evolve in a coupled manner (except
from small deviations around the $m_{\bf k} = \omega_\textrm{out}$
line), they fluctuate from $0$ to $1$ until they reach the
homogeneous zero-state.  We also observe that fluctuations in
$m_{(3,9)}$ are larger than in $m_{(2,1)}$, given that degree
distribution make the number of nodes in class $(2,1)$ larger
than in class $(3,9)$.

\section{Voter model with link update}

The same assumptions and procedures apply to the link-update voter
model and the invasion process. The link update (LU) dynamics selects
first a directed connection, so that the node at the tail will
always transmit its state to the neighbor at the head.

The microscopic dynamics of the link-update voter model is described by
\begin{equation}
s_u(t+dt)=\mu_u(dt)\xi_u + (1-\mu_u(dt))s_u(t), \label{micro1nodeLU}
\end{equation}
where as for the voter dynamics $\xi_u$ is given by Eq.~(\ref{xi}) and the binary variable $\mu_u(dt)$ for the selection of a link has a probability distribution
\begin{equation}
P(\mu_u(dt)) = k_{u,\textrm{in}} dt \delta_{\mu_u(dt),1} + (1-k_{u,\textrm{in}}
dt)\delta_{\mu_u(dt),0}.
\label{muLU}
\end{equation}
A factor $\lambda/(N\left<k_\textrm{in}\right>)$ has been
reabsorbed in the definition of $dt$. Proceeding as for the voter
model (we skip the details), we arrive to the equation for the
evolution of the relative densities $m_{\bf k}$ for the different
degree classes,
\begin{equation}
\frac{d m_{\bf k}(t)}{dt}=-k_\textrm{in}m_{\bf k}(t)+ k_\textrm{in}\sum_{\bf k'} P_\textrm{in}({\bf k'}|{\bf
k}) m_{\bf k'}(t).
\label{driftL}
\end{equation}
Regarding the stationary state, the same result as for the voter
model is found. The state-one relative densities behave again as
$m_{\bf k}(t)= \sum_{\bf k'} P_\textrm{in}({\bf k'}|{\bf k}) m_{\bf
k'}(t)$, so that all the relative densities are entangled through
topological correlations. We can once again prove, within the
heterogeneous mean field approach and for correlated strongly
connected components, that a conserved quantity of the form
$\omega=\sum_{\bf k} \varphi_{\bf k} m_{\bf k}(t)$ exists and is
defined by the eigenvector problem
\begin{equation}
\tilde{\varphi}_{\bf k}=\sum_{\bf k'}  P_\textrm{in}({\bf k}|{\bf k'})
\tilde{\varphi}_{\bf k'},
\label{varphiV2}
\end{equation}
where now $\tilde{\varphi}_{\bf k}=k_\textrm{in} \varphi_{\bf k}$.
In general, it is not possible to derive these coefficients
without further specifying the form of degree-degree correlations
in the network.

When two-point correlations are absent,
\begin{equation}
\frac{d m_{\bf k}(t)}{dt}=-k_\textrm{in}m_{\bf k}(t)+k_\textrm{in}\omega_\textrm{out}(t).
\label{driftLunc}
\end{equation}
In the stationary state, $m_{\bf k}=\omega_\textrm{out}(t) \mbox{  }
\forall {\bf k}$, but $\omega_\textrm{out}(t)$
is not a conserved quantity for the link update process as it was for the voter model. Instead,
the conserved quantity is
\begin{eqnarray}
\omega_\textrm{oi}&=&\left<\frac{k_\textrm{out}}{k_\textrm{in}} m_{\bf
k}(t)\right>/\left<\frac{k_\textrm{out}}{k_\textrm{in}} \right> \nonumber \\
&=&\sum_{\bf
k}\frac{k_\textrm{out}}{k_\textrm{in}}P({\bf k})m_{\bf k}(t)/\langle
\frac{k_\textrm{out}}{k_\textrm{in}}\rangle,
\label{consLU}
\end{eqnarray}
which follows from
Eq.~(\ref{driftLunc}).
Compare this expression with that for the total magnetization in uncorrelated undirected networks  $w=\omega=\left( \sum_{k}P(k)m_{k}(t) \right) / \langle k\rangle$ which corresponds to the conserved quantity for those structures~\cite{Suchecki:2005a}. The dependence of the conserved weighted
magnetization on the ratio between out- and in-degree for directed networks
highlights the fact that in LU it is important to have both a high
out-degree to be influential and at the same time to have a low
in-degree not to be too influenceable. Notice that the
ratio of the directed degrees is well defined since we are assuming that the network is
organized at the macroscopic scale into a SCC without peripheral
components all nodes having at least one incoming and one
outgoing link. Finally, in finite systems the probability of the state-one absorbing state is given by the conserved quantity, $\omega_\textrm{oi}=P_1$, and so fixed by the initial condition.

The derivation of how the state-one relative densities converge to
their stationary value in uncorrelated networks is more intricate than for the voter model, but we can make
use of a quasi-stationary approximation~\cite{Gardiner:2004} in
order to solve Eq.~(\ref{driftLunc}), exploiting the fact that
$\omega_\textrm{oi}$ is the conserved quantity. In the stationary
state $\omega_\textrm{out}=\omega_\textrm{oi}$, and we approximate
the equation by
\begin{equation}
\frac{d m_{\bf k}(t)}{dt}=-k_\textrm{in} m_{\bf k}(t)+k_\textrm{in} \omega_\textrm{oi}.
\end{equation}
For a given initial condition $m_{\bf k}(0)$, the solution is
\begin{equation}
m_{\bf k}(t)=\omega_\textrm{oi} (m_{\bf
k}(0)-\omega_\textrm{oi})e^{-k_\textrm{in}t}.
\end{equation}
As in the voter model, all the densities decay exponentially fast
to the stationary value $\omega_\textrm{oi}$, but in contrast not
all the densities decay with the same velocity, which depends on
the in-degree. Higher in-degree classes have smaller relaxation
times and decay faster than lower ones, but the transient is
always faster as compared to the VM.

\begin{table*}[t]
\caption{Conserved quantities for voter-like models in strongly connected components of directed
networks. 1st column, existence of conserved quantity for
correlated networks; 2nd column, conserved quantity for
uncorrelated networks; 3rd column, stationary values for the
relative densities; 4th column, density decay.}
\begin{tabular}{l|cccc}
& \hspace{0.3cm}$\omega_\textrm{corr}$ & $\omega_\textrm{unc}$
& $m_{\bf k}^{st}$ & $m_{\bf k}$ \\ \hline \hline \\
 \hspace{0.3cm}VM  \hspace{0.3cm} & $\exists$ & $\omega_\textrm{out}=\frac{1}{\langle k_\textrm{out} \rangle}\sum_{\bf k}k_\textrm{out}P({\bf k})m_{\bf
k}(t)$ & \hspace{0.5cm}$\omega_\textrm{out}$\hspace{0.5cm} & $m_{\bf k}^{st}+(m_{\bf k}(0)-m_{\bf
k}^{st})e^{-t}$\\ \\
 \hspace{0.3cm}LU \hspace{0.3cm} & $\exists$ & $\omega_\textrm{oi}=\frac{1}{\langle k_\textrm{out}/k_\textrm{in} \rangle}\sum_{\bf k}k_\textrm{out}/k_\textrm{in}P({\bf k})m_{\bf
k}(t)$ & $\omega_\textrm{oi}$ & $m_{\bf k}^{st}+(m_{\bf k}(0)-m_{\bf
k}^{st})e^{-k_\textrm{in}t}$\\ \\
 \hspace{0.3cm}IP \hspace{0.3cm} & $\exists$ & $\omega_\textrm{in}=\frac{1}{\langle 1/k_\textrm{in} \rangle}\sum_{\bf k}
\frac{1}{k_\textrm{in}}P({\bf k})m_{\bf k}(t)$ & $\omega_\textrm{in}$ & $m_{\bf k}^{st}+(m_{\bf
k}(0)-m_{\bf k}^{st})e^{-\frac{k_\textrm{in}}{\left<k_\textrm{in} \right>}t}$ \\ [-0.2 cm]\\
\hline\hline
\end{tabular}
\label{table1}
\end{table*}

\section{Invasion process}
The invasion process (IP) picks nodes at
random that export their state to a randomly chosen outgoing
neighbor. A certain node $u$ will update its state in a passive
form only when one of its incoming neighbors $v$ is selected as
the first node in one iteration of the dynamics and then $v$
chooses $u$ among all its outgoing neighbors to transmit it its
state. In this situation, it is more convenient to work with the
probability of node $u$ undergoing a state update with final state
1, $\xi_u^{(1)}$, and the probability of node $u$ undergoing a
state update with final state 0, $\xi_u^{(0)}$. The probability
distributions of these dichotomic stochastic variables are
\begin{eqnarray}
P(\xi_u^{(1)}) &=& \Phi_u^1 dt \delta_{\xi_u^{(1)},1} +
(1-\Phi_u^1 dt)\delta_{\xi_u^{(1)},0}, \\ P(\xi_u^{(0)}) &=&
\Phi_u^0 dt \delta_{\xi_u^{(0)},1} + (1-\Phi_u^0
dt)\delta_{\xi_u^{(0)},0},
\label{xiIP}
\end{eqnarray}
with
\begin{eqnarray}
\Phi_u^1(t) &=& \sum_v a_{vu}s_v(t)/k_{v,\textrm{out}},\\
\Phi_u^0(t)&=& \sum_v
a_{vu}(1-s_v(t))/k_{v,\textrm{out}}
\end{eqnarray}
and the parameter $\lambda$ of the Poisson process for the happening of events reabsorbed in $dt$.
Using these expressions, the dynamics is described at the microscopic scale by
\begin{eqnarray}
s_u(t+dt)&=&\xi_u^{(1)}(dt)(1-\xi_u^{(0)}(dt)) \nonumber \\ &+&
(1-\xi_u^{(1)}(dt))(1-\xi_u^{(0)}(dt))s_u(t). \label{micro1nodeIP}
\end{eqnarray}

Following the same methodology as for the voter model, the drift
equations for the relative densities in the different degree
classes read
\begin{equation}
\frac{d m_{\bf k}(t)}{dt}=k_\textrm{in}\sum_{\bf k'}\frac{1}{k'_\textrm{out}} P_\textrm{in}({\bf k'}|{\bf k})
(m_{\bf k'}(t)-m_{\bf k}(t)).
\label{driftIP}
\end{equation}
The existence of a conserved quantity $\omega=\sum_{\bf k}
\varphi_{\bf k} m_{\bf k}(t)$ in the correlated case is governed by
the eigenvalue problem
\begin{equation}
\tilde{\varphi}_{\bf k}=\sum_{\bf k'}  \frac{P_\textrm{in}({\bf k}|{\bf
k'})/k_\textrm{out}}{\sum_{\bf k''}P_\textrm{in}({\bf k''}|{\bf k'})/k_\textrm{out}''}\tilde{\varphi}_{\bf k'},
\end{equation}
where $\tilde{\varphi}_{\bf k}=\varphi_{\bf k} k_\textrm{in}
\sum_{\bf k''}P_\textrm{in}({\bf k''}|{\bf k})/k_\textrm{out}''$.
Summing both sides of this equation over ${\bf k}$, one arrives
once more to a trivial identity and so a conserved quantity exists
in general on networks with degree-degree correlations. As we see
next, we can be more specific on uncorrelated networks, for which
Eq.~(\ref{driftIP}) reduces to
\begin{equation}
\frac{d m_{\bf k}(t)}{dt}=\frac{k_\textrm{in}}{\left<k_\textrm{in}\right>}(m(t)-m_{\bf k}(t)),
\label{driftIPunc}
\end{equation}
where $m(t)=\sum_{\bf k} P({\bf k}) m_{\bf k}(t)$ is the total
density of state-one nodes in the network.

In the stationary state, $m_{\bf k}(t)= m(t) \mbox{  } \forall {\bf k}$,
but here $m(t)$ is not a conserved quantity for the IP in
uncorrelated directed networks. Instead, the conserved quantity is
\begin{eqnarray}
\omega_\textrm{in}(t)&=&\left< \frac{m_{\bf k}(t)}{k_\textrm{in}} \right> /
\left< \frac{1}{k_\textrm{in}} \right> \nonumber \\
&=&\sum_{\bf k}\frac{1}{k_\textrm{in}}P({\bf
k})m_{\bf k}(t)/\left<\frac{1}{k_\textrm{in}} \right>.
\end{eqnarray}
In finite systems, the probability of the state-one absorbing state is given by this conserved quantity, $\omega_\textrm{in}=P_1$, and is therefore fixed by the initial condition. The dependence of the weights on the inverse of the in degree implies that those
nodes with low in-degree, so less influenceable, have the highest
contribution and control the process. This dependence on the in degree is analogous to the dependence on the degree of the conserved quantity $w=\omega=\left( \sum_{k}1/k P(k)m_{k}(t) \right) / \langle k\rangle$ in uncorrelated undirected networks~\cite{Sood:2005}.

After a transient, $m(t)$ reaches the value
$\omega_\textrm{in}$, so that the stationary values
of the relative densities are $m_{\bf k}(t)=\omega_\textrm{in} \mbox{
} \forall {\bf k}$. This result tells us that all the densities
become independent of ${\bf k}$ and reach the same stationary
value, as in the previous processes.

The derivation of how the state-one relative densities converge to their
stationary value in uncorrelated networks is more
intricate than for the voter model, but like for the link update we can make
use of a quasi-stationary approximation~\cite{Gardiner:2004} in order to solve
Eq.~(\ref{driftIPunc}). Substituting into Eq.~(\ref{driftIPunc}) that in the stationary state $m(t)=\omega_\textrm{in}$,
\begin{equation}
\frac{d m_{\bf
k}(t)}{dt}=\frac{k_\textrm{in}}{\left<k_\textrm{in}\right>}\left(\omega_\textrm{in}-m_{\bf
k}(t)\right).
\end{equation}
For a given initial condition $m_{\bf k}(0)$, the solution is
\begin{equation}
m_{\bf k}(t)=\omega_\textrm{in} + \left(m_{\bf
k}(0)- \omega_\textrm{in} \right)e^{-\frac{k_\textrm{in}}{\left<k_\textrm{in}\right>}t}.
\end{equation}
All the densities decay exponentially to the stationary value
$\omega_\textrm{in}$. Higher in-degree classes decay faster than
lower ones with a relaxation time that is proportional to the
inverse of the in-degree, as is the case for LU. Due to
the average degree in the relaxation time, however, transients
are generally slower in the IP than in the LU. When compared with
the VM, the IP dynamics exhibits a slower transient for degree
classes with in-degree below average while those with in-degree
above the average converge faster to the stationary state.

\section{Conclusions}
We have introduced an analytical formalism from microscopic dynamics to show
that three different nonequilibrium dynamical models with two-absorbing states running on strongly
connected components of directed networks with heterogeneous degrees and
degree-degree correlations have associated ensemble average conservation
laws. These conservation laws have been fully determined when degree-degree correlations are
absent. The existence of ensemble average conservation laws is a general
characteristic of Markov processes with two or more absorbing states.

The constraints imposed
on the dynamics by the conservation laws lead to interesting and nontrivial
behavior. From a practical point of view, they are related to the stationary
values and the characteristic relaxation times of the relative densities of nodes in state one in each degree class and, in finite
systems, gives the
probabilities of reaching the two possible absorbing states. In this sense, the conservation laws obtained in the thermodynamic limit for a system that does not order in that limit (i.e. does not reach the absorbing state) determine the probabilities of reaching each absorbing state for a finite system. The contribution of each node
to he conserved global weighted magnetization is always a
specific function of the directed degrees. In the case of the VM, the
out-degree is the weight that controls the importance of the node as a measure
of its influence, while in the IP it is the inverse of the in-degree, and in
the LU it is the ratio between out and in-degree. In all cases, the conserved quantities
are determined by local properties that encode the importance of each node in the network. Depending on the dynamics,
what seems important from a local perspective is to be influential reaching a
large number of neighbors, or not to be too influenceable, with a low number
of incoming connections, or both at the same time.

From a broad perspective, these studies help in the understanding of how the rich structure of real systems
affects the dynamical processes that run on top. However, many questions still remain unsolved. In which specific way do degree
correlations alter the results for uncorrelated networks? How is the diffusive
fluctuations regime in SCCs of finite directed networks? Is the finite size scaling of consensus times
the same as in undirected networks? On the other hand, it seems realistic to
restrict to SCCs for a number of densely connected systems, like for instance
the world trade web~\cite{Serrano:2007b}, but in sparse directed networks the whole
structure of core and peripheral components should be taken into account. Numerical simulations in some specific model networks~\cite{MinPark:2006} show that the appearance of an input component seems to prevent the system, even if finite, from reaching an absorbing state for specific initial conditions. How does the complete structure of a directed network couples to the initial conditions of the dynamics to
induce the presence of zealots and how do they affect in quantitative terms
the behavior of the whole system still needs further research.

During the final completion of this work, we became aware of a recent preprint~\cite{Masuda:2008} discussing the fixation probabilities of
mutants for Voter-like dynamics on directed networks. Since there exists a direct relation between fixation probabilities of mutants and exit probabilities, and so conserved quantities, some of the results derived in that paper --without reference to conservation laws- concerning the dependence on the directed degrees are in correspondence to some of our results on uncorrelated strongly connected components.

\begin{acknowledgments}
We thank Mari\'{a}n Bogu\~{n}\'{a} for helpful discussions. We
acknowledge financial support from MCIN (Spain) and FEDER through
project FISICOS (FIS2007-60327); M.~A.~S. acknowledges support by DGES grant No.
FIS2007-66485-C02-01, K.~K. acknowledges financial
support from the Volkswagen Stiftung.
\end{acknowledgments}


\end{document}